\documentstyle[amssymb,prb,aps,preprint,floats]{revtex}

\input epsf

\begin{document}
\title{{\sc Density Perturbations in Multifield Inflationary Models }}
\author{V.F. Mukhanov$^a$ and Paul J. Steinhardt$^b$}
\address{$^a$Institut f\"{u}r Theoretische Physik, ETH Zurich, CH-9093 Zurich,\\
Switzerland}
\address{$^b$Department of Physics and Astronomy\\
University of Pennsylvania \\
Philadelphia, Pennsylvania 19104 USA}
\maketitle

\begin{abstract}
We derive a closed-form, analytical expression for the spectrum of
long-wavelength density perturbations in inflationary models with two (or
more) inflaton degrees of freedom that is valid in the slow-roll
approximation. We illustrate several classes of
potentials for which this expression reduces to a simple, algebraic
expression.
\end{abstract}

\baselineskip 25pt

\newpage

\section{Introduction}

Inflationary cosmology, in addition to resolving the cosmological flatness
and horizon problems of the standard big bang model,\cite{Guth} predicts a
remnant spectrum of density 
perturbations.\cite{MC,GP,Hawking,Starobinskii,BST,MFB} The perturbations may have seeded
large-scale structure formation and may have left an imprint on the cosmic
microwave background (CMB) anisotropy. Heretofore, considerable attention
has been given to precise derivations of the perturbation spectrum when a
single scalar (inflaton) field drives inflation and simultaneously this
field is responsible for origin of initial inhomogeneities.\cite
{GP,Hawking,Starobinskii,BST,MC,MFB}  Recently, we have shown how different
methods of computing the perturbation are related and compared their
accuracy.\cite{Wangetal} These results have been applied to obtain
predictions of CMB anisotropy and large-scale structure.

In this paper, we derive the explicit, analytical solution for the
perturbations in the models with two (or more) inflaton degrees of freedom
in a case of general potential for the scalar fields. To obtain this expression,
we must assume that the slow-roll approximation for the evolution of the scalar
fields is valid. With two or more
inflaton fields, the perturbation spectrum has additional contributions
which do not occur in the single-field case; in particular, the single field
spectrum consists purely of adiabatic fluctuations whereas the multi-field
spectrum generically has an entropic contribution as well, as discussed for
some specific cases previously.\cite{KL,MZ,Polarski,LM,G-BW,Yokoyama,Sal,SS}
Previously, a formalism has been developed for cases in which the two fields
are not coupled or have very simple couplings.\cite{Polarski,G-BW,Yokoyama,SS} 
However, in more complicated cases, only
heuristic arguments based on single-field inflation have been applied. Yet,
the single-field case is unusual because the evolving expectation value of
the one field serves as a clock that determines when the universe exits
inflation and returns to Friedmann-Robertson-Walker expansion. This was the
basis, for example, of the time-delay formalism introduced by Guth and Pi%
\cite{GP} for computing the perturbation spectrum for single-inflaton
models. With two or more fields, fluctuations in one field can affect the
evolution of the other field, and the complex conditions under which
inflation ends cannot be expressed in terms of one degree of freedom ({\it %
e.g.,} some linear combination of fields). Consequently, the heuristic
arguments are suspect. The issue has become more important in recent years
because intriguing new models of inflation have been proposed which entail
two or more inflaton degrees of freedom: {\it e.g.,} extended and
hyperextended inflation,\cite{extended,hyperext} hybrid inflation,\cite{hybrid} and supernatural inflation.\cite{supernat,supernat1}

For these reasons, it has become essential to have a formalism that applies
to the multifield case. Below we develop a procedure based on the natural
generalization for the single-field case. We assume the slow-roll
approximation in which the inflation kinetic energy is negligible compared
to its potential energy during inflation. We first review the single-field
case, then a simple case with two decoupled fields, and finally we solve the
equations for perturbations in the general case. The central result is in
Section IVb, Eqs.~(\ref{solp}) through~(\ref{solpot}), the closed-form
expressions for the perturbation spectrum. We present a series of potential
forms for which the closed-form expressions reduce to simple algebraic
expressions.

\section{Perturbations in Single-field inflation}

We consider first the case of a single scalar field $\phi $ with potential $%
V(\phi )$. Assuming the spatial part of the energy-momentum tensor is
diagonal, the metric in longitudinal gauge is: 
\begin{equation}
ds^{2}=(1+2\Phi )dt^{2}-a^{2}(1-2\Phi )\delta _{ik}dx^{i}dx^{k},
\end{equation}
where $a$ is the Robertson-Walker scale factor and $\Phi $ is the
gravitational potential. The perturbed Klein-Gordon equation which describes
the evolution of perturbations in $\phi $ is (in units with $\hbar =4\pi
G=c=1$): 
\begin{equation}
\ddot{\delta \phi }+3H\dot{\delta \phi }-\frac{1}{a^{2}}\nabla ^{2}\delta
\phi +V^{\prime \prime }\delta \phi -4\dot{\phi}\dot{\Phi}+2V^{\prime }\Phi
=0,
\end{equation}
and the perturbed $0-i$ Einstein equation is 
\begin{equation}
\dot{\Phi}+H\Phi =\dot{\phi}_{0}\delta \phi ,
\end{equation}
where $\phi =\phi _{0}+\delta \phi $, dot means $\partial /\partial t$, and
prime is used for $\partial /\partial \phi $. Throughout we use
dimensionless units where $4 \pi G = 1$, where $G$ is Newton's constant.

To find the nondecaying solution for the long-wavelength inhomogeneities
(for which the spatial derivatives in the equations of motion can be
neglected) in slowroll approximation, the terms proportional to $\dot{\Phi}$
or depending on second derivatives in $\delta \phi $ can be dropped,
resulting in the simplified equations 
\begin{equation}
3H\dot{\delta \phi }+V^{\prime \prime }\delta \phi +2V^{\prime }\Phi =0
\label{klein1}
\end{equation}
\begin{equation}
H\Phi =\dot{\phi}_{0}\delta \phi. \newline
  \label{klein2}
\end{equation}
The (unperturbed) background equations are: 
\begin{equation}
\begin{array}{rcl}
3H\dot{\phi}_{0} & = & -V^{\prime } \\ 
H^{2} & = & \frac{2}{3}V \\ 
\dot{H} & = & -\dot{\phi}_0^{2}
\end{array}
\label{six}
\end{equation}
where the slow-roll approximation has been assumed in dropping $\ddot{\phi}%
_{0}$ in the first equation and $\dot{\phi}^{2}$ in the second.

If we introduce a new variable, $x\equiv \delta \phi /V^{\prime }$, the
perturbative equations reduce to: 
\begin{equation}
\begin{array}{rcl}
3H\dot{x} & = & -2\Phi \\ 
\Phi & = & \frac{\dot{V}}{H}x.
\end{array}
\label{e1}
\end{equation}
Substituting the second expression into the first and using the background
equation to express $H$ in terms of $V$, we obtain 
\begin{equation}
\dot{x}=-\frac{\dot{V}}{V}x,
\end{equation}
whose solution is 
\begin{equation}
x=\frac{C}{V},
\end{equation}
where $C$ is an integration constant. Therefore, using the definition of $x$
and Eq.~(\ref{six}), we obtain 
\begin{equation}
\delta \phi =V^{\prime }x=C\frac{V^{\prime }}{V}=-2C\frac{\dot{\phi}_{0}}{H}.
\end{equation}
To fix the integration constant, we use the quantum de Sitter fluctuation
result that $\delta \phi _{k}\thicksim H$ evaluated at horizon-crossing, $%
k=aH$. Solving for $C$ above, we obtain 
\begin{equation}
C\approx \left( -H\frac{\delta \phi }{2\dot{\phi}}\right) _{k=aH}.
\label{eleven}
\end{equation}
According to Eq.~(\ref{e1}), $\Phi =\dot{V}x/H$. Since our solution is $%
x=C/V $, we obtain 
\begin{equation}
\Phi =C\frac{\dot{V}}{HV}=\,2C\frac{\dot{H}}{H^{2}}.  \label{single}
\end{equation}
The last expression is obtained by noting $H^{2}=\frac{2}{3}V$ and $\dot{H}%
/H=2\dot{V}/V$ where, applying the slow-roll approximation, we have ignored
the inflaton kinetic energy contribution to total energy density during
inflation. At the end of inflation, $\dot{H}={\cal O}(1)H^{2}$, which means
that 
\begin{equation}
\Phi _{f}\approx {\cal O}(1)C\approx {\cal O}(1)\left( H\frac{\delta \phi }{%
\dot{\phi}}\right) _{k=aH},  \label{final1}
\end{equation}
the standard, lowest-order result.\cite{Wangetal} The fluctuations described
in this relation are adiabatic. For the cosmic microwave
background (CMB) anisotropy in large angular scales, the gravitational
potential sets the CMB temperature fluctuations,\cite{SW} $\delta T/T\approx \Phi /3$.

\section{Two Decoupled Fields}

Next, consider the case of two decoupled fields $\phi _{1}$ and $\phi _{2}$
with potential $V(\phi _{1},\phi _{2})=V_{1}(\phi _{1})+V_{2}(\phi _{2})$.
This case has been considered previously.\cite{Polarski} In this case, the
Klein-Gordon equations for the fields in the slow-roll, long-wavelength
approximation takes the form: 
\begin{equation}
\begin{array}{rcl}
3H\dot{\delta \phi }_{1}+V_{1}^{\prime \prime }\delta \phi
_{1}+2V_{1}^{\prime }\Phi & = & 0 \\ 
3H\dot{\delta \phi }_{2}+V_{2}^{\prime \prime }\delta \phi
_{2}+2V_{2}^{\prime }\Phi & = & 0
\end{array}
\label{two1}
\end{equation}
and 
\begin{equation}
H\Phi =\dot{\phi}_{1}\delta \phi _{1}+\dot{\phi}_{2}\delta \phi _{2},
\label{two2}
\end{equation}
where $V_{1}^{\prime }\equiv \partial V_{1}/\partial \phi _{1}$ and $%
V_{2}^{\prime }\equiv \partial V_{2}/\partial \phi _{2}$.

The (unperturbed) background equations are: 
\begin{equation}
\begin{array}{rcl}
3H\dot{\phi}_{1} & = & -V_{1}^{\prime } \\ 
3H\dot{\phi}_{2} & = & -V_{2}^{\prime } \\ 
H^{2} & = & \frac{2}{3}(V_{1}+V_{2}) \\ 
\dot{H} & = & -(\dot{\phi}_{1}^{2}+\dot{\phi}_{2}^{2})
\end{array}
\end{equation}

Similar to the single-field case, we introduce two new variables $x_{1}$ and 
$x_{2}$ via 
\begin{equation}  \label{assign}
\delta \phi _{1}=V_{1}^{\prime }x_{1}\;\;{\rm and}\;\;\delta \phi
_{2}=V_{2}^{\prime }x_{2}.
\end{equation}
Then, the system of perturbed equations, Eq.~(\ref{two1}-\ref{two2}), can be
rewritten as 
\begin{equation}
3H\dot{x}_{1}=-2\Phi  \label{twoa}
\end{equation}
\begin{equation}
3H\dot{x}_{2}=-2\Phi  \label{twob}
\end{equation}
\begin{equation}
\Phi =\frac{1}{H}(\dot{V}_{1}x_{1}+\dot{V}_{2}x_{2}).  \label{twoc}
\end{equation}
Subtracting Eq.~(\ref{twob}) from Eq.~(\ref{twoa}), we obtain $\dot{x}_{1}-%
\dot{x}_{2}=0$, or 
\begin{equation}
x_{2}=x_{1}+D,
\end{equation}
where $D$ is an integration constant. Substituting this into Eq~(\ref{twoc}%
), one gets that 
\begin{equation}
\Phi =\frac{1}{H}[(\dot{V}_{1}+\dot{V}_{2})x_{1}+D\dot{V}_{2}].
\label{solPhi}
\end{equation}
Using this expression to replace the right-hand-side of Eq.~(\ref{twoa}), we
then obtain the closed form equation for $x_{1}$: 
\begin{equation}
\dot{x}_{1}=-\frac{\dot{V}_{1}+\dot{V}_{2}}{V_{1}+V_{2}}x_{1}-\frac{D\dot{V}%
_{2}}{V_{1}+V_{2}}
\end{equation}
(where the background equation expressing $H^{2}$ in terms of $V_{1}$ and $%
V_{2}$ has been used). This equation may be rewritten as 
\begin{equation}
\frac{\partial }{\partial t}(x_{1}(V_{1}+V_{2})+DV_{2})=0,
\end{equation}
and then can be integrated to obtain 
\begin{equation}
x_{1}=\frac{C-DV_{2}}{V_{1}+V_{2}},  \label{sol1}
\end{equation}
where $C$ is the integration constant. The corresponding result for $x_{2}$
is 
\begin{equation}
x_{2}=\frac{C+DV_{1}}{V_{1}+V_{2}}.  \label{sol2}
\end{equation}
The integration constants $C$ and $D$ should be fixed by assuming that $%
\delta \phi _{1,2}\sim H$ at horizon crossing during inflation, $k=aH$.
Using the relation between $x_{1,2}$ and $\delta \phi _{1,2}$, we obtain 
\[
D  =  \left[ \frac{1}{3H}\left( \frac{\delta \phi _{1}}{\dot{\phi}_{1}}-%
\frac{\delta \phi _{2}}{\dot{\phi}_{2}}\right) \right] _{k=aH} 
\]
\begin{equation}
C  =  -\frac{1}{2}\left[ H\left( \frac{V_{1}}{V_{1}+V_{2}}\frac{\delta
\phi _{1}}{\dot{\phi}_{1}}+\frac{V_{2}}{V_{1}+V_{2}}\frac{\delta \phi _{2}}{%
\dot{\phi}_{2}}\right) \right] _{k=aH}
\end{equation}
Also, substituting Eq.~(\ref{sol1}) into the expression for $\Phi $, Eq.~(%
\ref{solPhi}), we derive 
\begin{equation}
\Phi =2C\frac{\dot{H}}{H^{2}}+D\frac{1}{H}\frac{V_{1}\dot{V}_{2}-\dot{V}%
_{1}V_{2}}{V_{1}+V_{2}}
\end{equation}
The first term above is the same as the total contribution in the single
field case (see Eq.~(\ref{single})); as in the single-field case, it can be
interpreted as the adiabatic contribution to the fluctuation spectrum. The
second term is a new contribution due to entropic fluctuations which arises
whenever there are two or more inflaton fields. The entropic contribution
arises because two or more components of the cosmic fluid are undergoing
different fluctuations and evolution. Hence, the entropic contribution is a
characteristic feature of multicomponent inflation models.\cite{Polarski}

\section{General Case}

We now generalize our method to models with two scalar fields in which the
kinetic energy terms have non-linear
sigma-model form and the potential $V\left( \phi
,\psi \right) $ is an arbitrary function of $\phi ,\psi :$%
\begin{equation}
S=\int \left[ \frac{1}{2}f_{1}\left( \phi ,\psi \right) \phi _{;\alpha }\phi
^{;\alpha }+\frac{1}{2}f_{2}\left( \phi ,\psi \right) \psi _{;\alpha }\psi
^{;\alpha }-V\left( \phi ,\psi \right) \right] \sqrt{-g}d^{4}x
\end{equation}
Kinetic terms of this type occur, for example, in supergravity models with
non-trivial Kahler potential.

\subsection{Derivation}

Variation of this action with respect to $\phi ,\psi $ leads to the
equations 
\begin{equation}
\left( f_{1}\phi ^{;\alpha }\right) _{;\alpha }-\frac{1}{2}f_{1,\phi }\phi
_{;\alpha }\phi ^{;\alpha }-\frac{1}{2}f_{2,\phi }\psi _{;\alpha }\psi
^{;\alpha }+V_{,\phi }=0
\end{equation}
\begin{equation}
\left( f_{2}\psi ^{;\alpha }\right) _{;\alpha }-\frac{1}{2}f_{1,\psi }\phi
_{;\alpha }\phi ^{;\alpha }-\frac{1}{2}f_{2,\psi }\psi _{;\alpha }\psi
^{;\alpha }+V_{,\psi }=0
\end{equation}
We consider a Friedmann-Robertson-Walker Universe with small perturbations 
\begin{equation}
ds^{2}=\left( 1+2\Phi \right) dt^{2}-a^{2}\left( 1-2\Phi \right) \delta
_{ik}dx^{i}dx^{k}
\end{equation}
and decompose fields into homogeneous and inhomogeneous components 
\begin{equation}
\phi =\phi _{0}\left( t\right) +\delta \phi \left( x,t\right) ,\psi =\psi
_{0}+\delta \psi
\end{equation}
The equation for the background then take the form 
\begin{equation}
\left( \ddot{\phi}_{0}+3H\dot{\phi}_{0}\right) f_{1}+\dot{f}_{1}\dot{\phi}%
_{0}-\frac{1}{2}f_{1,\phi }\dot{\phi}_{0}^{2}-\frac{1}{2}f_{2,\phi }\dot{\psi%
}_{0}^{2}+V_{,\phi }=0
\end{equation}
Here all functions depend only on the background variables. The equation for 
$\psi $ can be obtained from the above one if we make the following
substitutions: $f_{1}\Leftrightarrow f_{2}$; $\phi \Leftrightarrow \psi .$

The first linearized equation for perturbations is 
\[
f_{1}(\delta \ddot{\phi}+3H\delta \dot{\phi}-\frac{1}{a^{2}}\nabla
^{2}\delta \phi -4\dot{\Phi}\dot{\phi}_{0})+2\Phi V_{,\phi }+f_{1,\phi }\dot{%
\phi}_{0}\delta \dot{\phi}-f_{2,\phi }\dot{\psi}_{0}\delta \dot{\psi}%
+f_{1,\psi }\left( \dot{\phi}_{0}\delta \dot{\psi}+\dot{\psi}_{0}\delta \dot{%
\phi}\right) 
\]
\[
\left[ f_{1,\phi }\left( \ddot{\phi}_{0}+3H\dot{\phi}_{0}\right) +\frac{1}{2}%
f_{1,\phi \phi }\dot{\phi}_{0}^{2}-\frac{1}{2}f_{2,\phi \phi }\dot{\psi}%
_{0}^{2}+f_{1,\phi \psi }\dot{\phi}_{0}\dot{\psi}_{0}+V_{,\phi \phi }\right]
\delta \phi + 
\]
\begin{equation}
\left[ f_{1,\psi }\left( \ddot{\phi}_{0}+3H\dot{\phi}_{0}\right) +\frac{1}{2}%
f_{1,\phi \psi }\dot{\phi}_{0}^{2}-\frac{1}{2}f_{2,\phi \psi }\dot{\psi}%
_{0}^{2}+f_{1,\psi \psi }\dot{\phi}_{0}\dot{\psi}_{0}+V_{,\phi \psi }\right]
\delta \psi =0
\end{equation}
The second equation is obtained by the above mentioned substitution. Because
we have three unknown functions, we need a third relation. It is convenient
to choose the 0-$i$ Einstein equation: 
\begin{equation}
\dot{\Phi}+H\Phi =f_{1}\dot{\phi}_{0}\delta \phi +f_{2}\dot{\psi}_{0}\delta
\psi
\end{equation}

The above equations can be simplified and integrated explicitly for the
longwave perturbations in slowroll approximation. In this approximation the
equations for the background simplifies to 
\[
3Hf_{1}\dot{\phi}_{0}=-V_{,\phi } 
\]
\[
3Hf_{2}\dot{\psi}_{0}=-V_{,\psi } 
\]
\begin{equation}
H^{2}=\frac{2}{3}V  \label{back}
\end{equation}
and the equations for perturbations take the form 
\[
3H\delta \dot{\phi}+\left( \frac{V_{,\phi }}{f_{1}}\right) _{,\phi }\delta
\phi +\left( \frac{V_{,\phi }}{f_{1}}\right) _{,\psi }\delta \psi +2\Phi 
\frac{V_{,\phi }}{f_{1}}=0 
\]
\[
3H\delta \dot{\psi}+\left( \frac{V_{,\psi }}{f_{2}}\right) _{,\phi }\delta
\phi +\left( \frac{V_{,\psi }}{f_{2}}\right) _{,\psi }\delta \psi +2\Phi 
\frac{V_{,\psi }}{f_{2}}=0 
\]
\begin{equation}
H\Phi =f_{1}\dot{\phi}_{0}\delta \phi +f_{2}\dot{\psi}_{0}\delta \psi
\end{equation}
Let us define the variables \ $x$ and $y$ as 
\begin{equation}
\delta \phi =\frac{V_{,\phi }}{f_{1}}x\,\,{\rm and}\,\,\,\,\,\,\,\,\,\,\,\,%
\,\delta \psi =\frac{V_{,\psi }}{f_{2}}y.  \label{defin}
\end{equation}
In terms of these variables, the above equations-of-motion for the
perturbations become 
\[
3H\dot{x}+\frac{\left( V_{,\phi }/f_{1}\right) _{,\psi }\left( V_{,\psi
}/f_{2}\right) }{\left( V_{,\phi }/f_{1}\right) }\left( y-x\right) +2\Phi =0 
\]
\begin{equation}
3H\dot{y}+\frac{\left( V_{,\psi }/f_{2}\right) _{,\phi }\left( V_{,\phi
}/f_{1}\right) }{\left( V_{,\psi }/f_{2}\right) }\left( x-y\right) +2\Phi =0
\label{fortyone}
\end{equation}
These equations are very similar to Eqs.~(\ref{twoa}) and (\ref{twob}) for
the decoupled case, except that there is here the middle term which vanishes
if $V=V_{1}(\phi )+V_{2}(\psi )$. Nevertheless, we can obtain a closed-form
solution. Subtracting the two equations, we obtain 
\begin{equation}
3H\left( \dot{y}-\dot{x}\right) =\left( \frac{\left( V_{,\phi }/f_{1}\right) _{,\psi
}\left( V_{,\psi }/f_{2}\right) }{\left( V_{,\phi }/f_{1}\right) }+\frac{%
\left( V_{,\psi }/f_{2}\right) _{,\phi }\left( V_{,\phi }/f_{1}\right) }{%
\left( V_{,\psi }/f_{2}\right) }\right) \left( y-x\right)
\end{equation}
which can be integrated 
\begin{equation}  \label{fortythree}
y-x= \gamma \exp \left\{ \int \left[ \frac{\left( V_{,\phi }/f_{1}\right)
_{,\psi }\left( V_{,\psi }/f_{2}\right) }{\left( V_{,\phi }/f_{1}\right) }+%
\frac{\left( V_{,\psi }/f_{2}\right) _{,\phi }\left( V_{,\phi }/f_{1}\right) 
}{\left( V_{,\psi }/f_{2}\right) }\right] \frac{dt}{3H}\right\},
\label{fint}
\end{equation}
where $\gamma$ is a constant of integration. Using the $0-i$ Einstein
equation, $\Phi $ can be expressed in terms of $x$%
\begin{equation}
\Phi =\frac{1}{H}\left( f_{1}\dot{\phi}\delta \phi +f_{2}\dot{\psi}\delta
\psi \right) =\frac{1}{H}\left( V,_{\phi }\dot{\phi}x+V,_{\psi }\dot{\psi}%
y\right) =\frac{1}{H}\left( \dot{V}x+V,_{\psi }\dot{\psi}\left( y-x\right)
\right)  \label{phi}
\end{equation}
Eqs.~(\ref{fortyone}) and (\ref{fortythree}) can then be used to find an
integral expression for $x$: 
\begin{equation}
x=-\frac{\gamma}{V}\int \left[ H\frac{\left( V_{,\phi }/f_{1}\right) _{,\psi
}\left( V_{,\psi }/f_{2}\right) }{2\left( V_{,\phi }/f_{1}\right) }+V,_{\psi
}\dot{\psi}\right] \frac{F}{V}dt  \label{first}
\end{equation}
where 
\begin{equation}
F=V\exp \left\{ \int \left[ \frac{\left( V_{,\phi }/f_{1}\right) _{,\psi
}\left( V_{,\psi }/f_{2}\right) }{\left( V_{,\phi }/f_{1}\right) }+\frac{%
\left( V_{,\psi }/f_{2}\right) _{,\phi }\left( V_{,\phi }/f_{1}\right) }{%
\left( V_{,\psi }/f_{2}\right) }\right] \frac{dt}{3H}\right\}  \label{def1}
\end{equation}
Similar expressions can be obtained for $y.$ Using the definitions of $x$
and $y$ (Eq.~(\ref{defin}) and the equations for the background (\ref{back}%
), the expressions can be simplified to the final, compact,  closed forms 
given in the next section.

\subsection{A General, Closed-form Expression}

The following closed-form expressions  are a compact, general representation of the perturbations  with wavenumber $k$ 
in two-field models in the slow-roll approximation, the central result
of this paper.  The perturbed variables
$\delta \phi$, $\delta \psi$, and $\Phi$ are  $k$-dependent, but the 
subscript $k$ has been suppressed for simplicity.
\begin{equation}
\delta \phi (t)=\gamma \frac{\left( \ln V\right) _{,\phi }}{f_{1}}%
\int_{t_{0}}^{t}\left( \ln \left( \frac{\left( \ln V\right) ,_{\phi }}{f_{1}}%
\right) \right) _{,\psi }Fd\psi + \alpha \frac{\left( \ln V\right) _{,\phi }%
}{f_{1}}  \label{solp}
\end{equation}
\begin{equation}
\delta \psi (t)=-\gamma \frac{\left( \ln V\right) _{,\psi }}{f_{2}}%
\int_{t_{0}}^{t}\left( \ln \left( \frac{\left( \ln V\right) ,_{\psi }}{f_{2}}%
\right) \right) _{,\phi }Fd\phi +\beta \frac{\left( \ln V\right) _{,\psi }}{%
f_{2}}  \label{solc}
\end{equation}
where 
\begin{equation}
F(t)=\exp \left\{ -\int_{t_{0}}^{t}\left[ \left( \ln \left( \frac{\left( \ln
V\right) ,_{\psi }}{f_{2}}\right) \right) _{,\phi }d\phi +\left( \ln \left( 
\frac{\left( \ln V\right) ,_{\phi }}{f_{1}}\right) \right) _{,\psi }d\psi
\right] \right\}  \label{def2}
\end{equation}
(In re-expressing $F$, an overall constant has been removed and absorbed
into the definition of $\gamma$.)
Here $t_{0}$ is an arbitrary moment of time; we take $t_0$ to be the moment
of horizon crossing $t_{k}$ when $k=aH$ for the given mode $k$.
The limits of integration indicate that the integration variable is 
to be evaluated at the time $t_0$ and $t$.

There are three integrations constants, $\alpha$, $\beta$ and $\gamma$.
Evaluating these expressions at horizon-crossing and using the fact that $%
\delta \phi \sim \delta \psi \sim H$ at horizon-crossing, the integration
constants $\alpha$ and $\beta$ can be determined in terms of the inflaton
potential and derivatives during inflation. The third integration constant, $%
\gamma$, must be taken in such a way as to satisfy Eq.~(\ref{fint}), from
where it follows that $\beta=\alpha+ \gamma$. 
(One must make use of the definitions of $x$ and $y$ in Eq.~(\ref{defin})
and the fact that, at horizon-crossing, the integral on the right-hand-side of
Eq.~(\ref{fint}) is $1/V_{k=aH}$.)   
The value of $\gamma$
corresponds physically to the amplitude of the entropy perturbations. The
gravitational potential is expressed in terms of $\delta \phi $ and $\delta
\psi $ as 
\begin{equation}
\Phi =-\frac{1}{2}\left[ \left( \ln V\right) _{,\phi }\delta \phi +\left(
\ln V\right) _{,\psi }\delta \psi \right]  \label{solpot}
\end{equation}

The above formulae can also be rewritten in other forms which can be useful
for approximate evaluation of the integrals. For instance, in the standard
case where the kinetic energy terms are canonical ($f_{1}=f_{2}=1$), the
expressions above reduce to: 
\[
\delta \phi =-\gamma \left( \ln V\right) _{,\phi }\int_{t_{0}}^{t}\frac{%
V_{,\psi }^{2}}{V_{,\psi }^{2}+V_{,\phi }^{2}}dF+\alpha \left( \ln V\right)
_{,\phi } 
\]
\begin{equation}
\delta \psi =\gamma \left( \ln V\right) _{,\phi }\int_{t_{0}}^{t}\frac{%
V_{,\phi }^{2}}{V_{,\psi }^{2}+V_{,\phi }^{2}}dF+\beta \left( \ln V\right)
_{,\psi }
\end{equation}

\section{Applications}

In a surprisingly wide range of cases, the closed-form integral expressions
above can be simplified. Our first two examples are simple cases which have
been studied previously in the literature. We use these to show how our
general expressions are to be used and to test that they reproduce known
results. We then apply the method to more general and more realistic models.

\subsubsection{Example 1: \thinspace \thinspace \thinspace $%
f_{1}=f_{2}=1,\,\,\,\,\,\,V=V_{1}\left( \phi \right) + V_{2}\left( \psi
\right) $}

We have already presented a derivation for the most trivial case where the
two inflaton fields are decoupled, $V=V_{1}\left( \phi \right) +V_{2}(\psi )$
(see Section III). Here we show that the answer can be reproduced by our
more general formulae. We detail a few steps to aid the reader in becoming
familiar with applying our general formulae.

If $V=V_{1}(\phi )+V_{2}(\psi )$, then the expression for $F(t)$ in Eq.~(\ref
{def2}) can be simplified. The first integrand in the exponent can be
reduced to $-(\ln V)_{,\phi }d\phi $ and the second integrand to $-(\ln
V)_{,\psi }d\psi $. The two can be combined into a total differential, $%
-d(\ln V)$; as a result, $F$ reduces to $F(t)=V/V_{k=Ha}$, where $V_{k=Ha}$
is the value of potential taken at the moment of horizon crossing $%
t_{0}=t_{k}.$ In the first term of Eq.~(\ref{solp}), the integral reduces to 
$-\int_{t_{k}}^{t}(V_{2}^{^{\prime }}/V_{k=Ha})d\psi
=(V_{2}/V)_{k=Ha}-(V_{2}/V_{k=Ha})$ sothat 
\begin{equation}
\delta \phi =-\frac{\gamma }{V_{k=Ha}}\frac{V_{1}^{\prime }V_{2}}{V}+\left(
\alpha +\gamma \left( \frac{V_{2}}{V}\right) _{k=Ha}\right) \frac{%
V_{1}^{\prime }}{V},
\end{equation}
and 
\[
\delta \psi =\frac{\gamma }{V_{k=Ha}}\frac{V_{2}^{\prime }V_{1}}{V}+\left(
\beta -\gamma \left( \frac{V_{1}}{V}\right) _{k=Ha}\right) \frac{%
V_{2}^{\prime }}{V}
\]
which agrees with Eqs.~(\ref{assign}) and~(\ref{sol1}) in our earlier
derivation. Comparing these formulae with Eqs.~(\ref{assign}) and~(\ref{sol1}%
) we see that the integration constants $C$ and $D$ in Eq. (\ref{sol1}) are
linear combinations of $\gamma ,\alpha $ and $\beta $ : 
\[
C=\alpha +\gamma \left( \frac{V_{2}}{V}\right) _{k=Ha}=\beta -\gamma \left( 
\frac{V_{1}}{V}\right) _{k=Ha},
\]
\begin{equation}
D=\frac{\gamma }{V_{k=Ha}}.  \label{coeff}
\end{equation}

This identification between the three integration constants and the two
coefficients, $C$ and $D$, requires use of the constraint that\thinspace
\thinspace  $\alpha $ and $\beta $ must satisfy, $\beta -\alpha =\gamma $ ;
see remarks above Eq. (\ref{solpot}) in Section IVB. Also note that the
integration constants are functions of the wavenumber $k$ which can be
related to  $\delta \phi $ and $\delta \psi $ at the moment of horizon
crossing $k=Ha$ using the relations (\ref{solp}), (\ref{solc})  and taking
in these formulae $t=t_{k}:$ 
\[
\alpha =-\frac{1}{2}\left( H\frac{\delta \phi }{\dot{\phi}}\right)
_{k=Ha},\,\,\,\,\,\,\,\beta =-\frac{1}{2}\left( H\frac{\delta \psi }{\dot{%
\psi}}\right) _{k=Ha}
\]
\begin{equation}
\gamma =\frac{1}{2}\left[ H\left( \frac{\delta \phi }{\dot{\phi}}-\frac{%
\delta \psi }{\dot{\psi}}\right) \right] _{k=Ha}  \label{coe}
\end{equation}
Note that the above expressions for coefficients are valid in general case.

Taking into account the above relations (\ref{coeff}) we see that these
expressions for the constants of integration are in agreement with formulae
(27). We refer to this as Example 1. Other examples where the expressions
for the solutions can be significantly simplified are:

\subsubsection{Example 2: \thinspace \thinspace \thinspace $%
f_{1}=f_{2}=1,\,\,\,\,\,\,V=V_{1}\left( \phi \right) V_{2}\left( \psi
\right) $}

The integrands in the expression for $F(t)$ in Eq.~(\ref{def2}) and in the
expressions for $\delta \phi $ and $\psi $ are precisely zero. Hence, $F(t)=1
$ and 
\begin{equation}
\delta \phi =\alpha \frac{V_{1,\phi }}{V_{1}},\,\,\,\,\,\,\,\,\,\,\,\,\,%
\delta \psi =\beta \frac{V_{2,\psi }}{V_{2}}.
\end{equation}
Eq.~(\ref{solpot}) then reduces to a closed-form expression for $\Phi $, 
\begin{equation}
\Phi =-\frac{1}{2}\left( \alpha \frac{V_{1,\phi }^{2}}{V_{1}^{2}}+\beta 
\frac{V_{2,\psi }^{2}}{V_{2}^{2}}\right) ,
\end{equation}
in agreement with results obtained previously\cite{G-BW} for this kind of
potential.

\subsubsection{Example 3: $f_{1}=f_{1}\left( \psi \right)
,\,\,\,\,\,f_{2}=f_{2}\left( \phi \right) ,\,\,\,\,V=V_{1}(\phi )V_{2}\left(
\psi \right) $}

The solution is 
\[
\delta \phi =-\frac{\gamma }{\left( f_{1}f_{2}\right) _{k=Ha}}\frac{%
V_{1,\phi }}{V_{1}f_{1}}\int_{t_{k}}^{t}f_{2}df_{1}+\alpha \frac{V_{1,\phi }%
}{V_{1}f_{1}}
\]
\[
\delta \psi =\frac{\gamma }{\left( f_{1}f_{2}\right) _{k=Ha}}\frac{V_{2,\psi
}}{V_{2}f_{2}}\int_{t_{k}}^{t}f_{1}df_{2}+\beta \frac{V_{2,\psi }}{V_{2}f_{2}%
}
\]
where $\alpha ,\beta $ and $\gamma $ are given by the formulae (\ref{coe}).
In particular, when $f_{1}=\exp \left( -\psi /\psi _{1}\right)
,\,\,\,\,f_{2}=1,\,\,\,\,\,V_{2}\left( \psi \right) =\exp \left( -\psi /\psi
_{2}\right) $ we obtain 
\[
\delta \phi =\frac{V_{1,\phi }}{V_{1}}\left( \beta \exp \left( \psi /\psi
_{1}\right) -\gamma \exp (\psi _{k=Ha}/\psi _{1})\right) 
\]
\begin{equation}
\delta \psi =-\beta /\psi _{2}
\end{equation}
in  agreement with the recent result obtained by Starobinskii \&
Yokoyama.\cite{Yokoyama}

\subsubsection{Example 4: $f_{1}=f_{2}=1,\,\,\,\,\,\,V=\Lambda +V_{1}\left(
\phi \right) V_{2}\left( \psi \right) $}

In this case we have 
\[
\delta \phi =-\Lambda \gamma \left( 1-\frac{\Lambda }{V_{k=Ha}}\right)
\left( \ln V\right) _{,\phi }\int_{t_{k}}^{t}\frac{1}{V_{1}}d\left( \frac{1}{%
V_{2}}\right) +\alpha \left( \ln V\right) _{,\phi }
\]
\begin{equation}
\delta \psi =\Lambda \gamma \left( 1-\frac{\Lambda }{V_{k=Ha}}\right) \left(
\ln V\right) _{,\psi }\int_{t_{k}}^{t}\frac{1}{V_{2}}d\left( \frac{1}{V_{1}}%
\right) +\beta \left( \ln V\right) _{,\psi }
\end{equation}
where as before the coefficients $\alpha ,\beta $ and $\gamma $ can be
expressed in terms of $\delta \phi $ and $\delta \psi $ at the momemt of
horizon crossing via (\ref{coe}).

\subsubsection{Example 5: $f_{1}=f_{2}=1,\,\,\,\,\,\,V=\Lambda +\alpha
V_{1}\left( \phi \right) +\beta V_{2}\left( \psi \right) +V_{1}\left( \phi
\right) V_{2}\left( \psi \right) $}

This case can be reduced to the previous one by the redefinition of the
potentials $V_{1}$ and $V_{2}$ (shifting them by constant terms).

In summary, the central results of this paper are in Section IVb, explicit
analytical solutions of equations for the energy density perturbations in
the general case of two scalar fields with an arbitrary potential. To apply
these solutions to a particular inflationary model, one should find first the
analytical solutions which describe the behavior of the background and then
use the obtained solutions in the formulae we have provided. 
A significant feature compared
to the single inflaton field case is the existence of two constants of
integration, each set by conditions at horizon-crossing during inflation.
The two coefficients can be chosen so as to correspond to adiabatic and
entropic perturbations. The latter is absent in the single-field case, but
here is shown to be a general feature. In a future publication, we shall
discuss the observational consequences of these results; namely, how
observations can be used to distinguish single- from multi-field
inflationary models.

We thank G. Huey for suggestions for improving the manuscript.
This research was supported by the Department of Energy at Penn,
DE-FG02-95ER40893 (PJS), and by the Tomalla Foundation (VM).

\end{document}